\documentclass[a4paper,numerical,english,twocolumn,showpacs,prb]{revtex4-1}

\usepackage{amsmath, amsthm, amsfonts}
\usepackage{graphicx}
\usepackage{color}
\usepackage{natbib}
\usepackage[caption=false]{subfig}
\usepackage[colorlinks,linkcolor=blue,
            urlcolor=blue,
            citecolor=blue]{hyperref}
 
\newcommand{\eff}{\mathrm{eff}}

\newcommand{\mr}[1]{\mathrm{ #1}}

\begin{document}

\author{Sebastian Angst} 
\email[]{sebastian.angst@uni-due.de}
\author{Dietrich E. Wolf}
\affiliation{Faculty of Physics and CENIDE, University of
  Duisburg-Essen, D-47048, Duisburg, Germany}

\title{Network theory for inhomogeneous thermoelectrics}

\begin{abstract}
  The Onsager-de Groot-Callen transport theory, implemented as a
  network model, is used to simulate the transient Harman method,
  which is widely used experimentally to determine all thermoelectric
  transport coefficients in a single measurement setup. It is shown
  that this method systematically overestimates the Seebeck coefficient 
  for samples composed of two different materials.
  As a consequence, the figure of merit is also
  overestimated, if the thermal coupling of the measurement setup to
  the environment is weak. For a mixture of metal and semiconductor
  particles near metal percolation the figure of merit obtained by the
  Harman method is more than 100 \% too large. For a correct
  interpretation of the experimental data,
  information on composition and microstructure of the sample are indispensable.
\end{abstract}


\maketitle

\section{Introduction}

Thermoelectric materials are important for energy harvesting,
especially from waste heat \cite{Snyder2008,Shakouri2011}. In order to
optimize the conversion into electricity, it is desirable to predict
device properties by efficient computer simulations. This is one goal
of the present paper. More fundamentally, we are going to point out
that memory effects render the global response of composite materials,
which are common among modern nanostructured thermoelectrics
\cite{Dresselhaus2007}, highly complex.

The theoretical description of transport processes goes back to
Onsager \cite{Onsager1931,Onsager1931_2}. Applied to thermoelectrics,
this became known as the Onsager-de Groot-Callen theory
\cite{Callen1948}. A special case is the so-called constant property
model (CPM), where the electric and heat conductivity, $\sigma$ and
$\kappa$,  and the Seebeck coefficient $\alpha$ are
assumed to be constant. This
model has been studied in detail analytically in one dimension
\cite{Seifert2006,Goupil2011} and will serve as a reference system for
validation in this paper.

Since analytic calculations are restricted to simple compounds,
numerical models have been developed in order to describe
inhomogeneous materials. Although these models are also based on the
Onsager-de Groot-Callen theory, most of them do not fully describe the
thermoelectric effects, since they do not include Joule heat and/or
Peltier heat \cite{Webman1977, Kleber2005}. They were used to
calculate either the heat and electrical conductances or the Seebeck
coefficient \cite{Gather2011,Gather2011_2,Becker2012}.  More complex
models \cite{Wachutka1990} exist in the framework of drift-diffusion
models, and they are applied e.g. for the simulation of generators in
complex geometries \cite{Span2007}.

In this paper a simple and coherent way to discretize the Onsager
transport theory for thermoelectric materials will be used, which
includes all relevant effects and time-dependencies. It can be seen as
a version of the finite difference method reviewed in
\cite{Hogan2006}.  Originally it was designed for the investigation of
transport processes in nano-particle configurations
(\cite{Hartner2012},\cite{Becker2012},\cite{Angst2013}), which were
mapped to a network model. 
In this paper it will be applied to disordered bulk systems
rather than particle agglomerates.

The model, derived in Sec.~\ref{sec:model}, enables us to study
thermoelectric effects in geometries, which can not be solved
analytically. It is validated by comparing simulations and analytical
results for one-dimensional segmented CPM thermoelectrics.  As an
application, the transient Harman method is simulated focusing on
inhomogeneous systems like segmented thermoelectrics, superlattices
and composite media. Furthermore, we support and extend our findings
concerning segmented structures by an analytical treatment.

\section{Model}
\label{sec:model}

The Onsager-de Groot-Callen theory \cite{Goupil2011} describes the
flow of electrical current density $\bf j$ and heat current density
$\bf j_q$
\begin{align}
  {\bf j} &= -\sigma \left(\nabla \mu/q + \alpha \nabla T\right)
  \label{eq:j}\\
  {\bf j_q}
  &= - \kappa \nabla T + \Pi \ {\bf j} \label{eq:j_q}
\end{align}
driven by the gradients of temperature $T$ and electrochemical
potential $\mu$ ($q$ is the charge of the mobile particles). According to
the Kelvin relation the Peltier coefficient $\Pi$ is related to the
Seebeck coefficient $\alpha$ by $\Pi = \alpha T$. If the Seebeck
coefficient is zero, heat and electric transport decouple resulting in
Ohm's and Fourier's laws with electrical conductivity $\sigma$ and
heat conductivity $\kappa$.  Volumetric heat production due to Joule
heating taken into account by
\begin{equation}
  c \dot T + \nabla \cdot {\bf j_q} =  -\frac{\nabla \mu}{q}  \cdot {\bf j }
  \label{eq:div_heat},
\end{equation}
with heat capacity per volume, $c$.

Discretization of these equations in form of a network model is
achieved by assigning variable temperatures $T_i$ and electrochemical
potentials $\mu_i$ to each lattice site $i$. The bonds between
neighboring sites are characterized by an electric conductance
$G_{ij}$, a heat conductance $K_{ij}$, a Seebeck and a Peltier
coefficient, $\alpha_{ij}$ and $\Pi_{ij}$. In the network model the
electric current $I_{ij}$ and the heat current $I_{{\text q},ij}$
between sites $i$ and $j$ read
\begin{eqnarray}
  I_{ij} &=& G_{ij}\left(\frac{\mu_i-\mu_j}{q} + \alpha_{ij} (T_i -
    T_j)\right), \label{eq:el_curr_model}\\
  I_{{\text q},ij} &=& K_{ij} (T_i - T_j) + \Pi_{ij} I_{ij}\label{eq:heat_curr_model}.
\end{eqnarray}

The coefficients $G_{ij},K_{ij},\alpha_{ij},\Pi_{ij}$ depend on
the material properties at sites $i$ and $j$. Let $G_i$ denote the
conductance of the material site $i$. The electrical resistance
$1/G_{ij}$ of the bond between sites $i$ and $j$, half of which is
material $i$ or $j$, respectively, is modeled as
\begin{equation}
\frac{1}{G_{ij}}= \frac{1}{2G_i} + \frac{1}{2G_j} + R
\end{equation}
with an extra interface contribution $R$ \cite{Gather2011_2}. 
Likewise, the thermal conductance $K_{ij}$ is calculated from 
\begin{equation}
\frac{1}{K_{ij}}= \frac{1}{2K_i} + \frac{1}{2K_j} +
R_{\text q}.
\end{equation}
Being contact rather than material properties, the
interface contributions $R$ and $R_{\text q}$ are particularly
important, if one has only one 
crystalline material that is divided into grains by grain
boundaries. In this paper, however, we want to focus on compound materials,
where the different bulk properties have a strong influence on the
thermoelectric response. In order not to obscure this bulk effect by
the additional influence of the interfaces, we set $R=0$ and $R_{\text q}=0$ in
the following.

Next, we discuss the Seebeck coefficients $\alpha_{ij}$, if the
materials at the neighboring sites $i$ and $j$ are different. The
Seebeck voltage $\alpha_{ij} (T_i - T_j)$ is the sum of two
contributions: first, the voltage between site $i$ and the interface
in the middle of the bond $ij$, which is at temperature $T_{c,ij}$,
and second, the voltage between the interface and site $j$:
\begin{align}
  &\alpha_{ij} \left(T_i - T_j\right) = \alpha_i \left(T_i -T_{c,ij} \right) + \alpha_j \left(T_{c,ij} -T_j \right)\nonumber \\ 
 =& \bar \alpha_{ij}\left( T_i - T_j \right) + \Delta \alpha_{ij}  \left(\bar T_{ij}-T_{c,ij}\right)\label{eq:seebeck_term},
\end{align}
where $\bar \alpha_{ij} =(\alpha_i + \alpha_j)/2$, $\bar T_{ij} =(T_i+
T_j)/2$ and $\Delta \alpha_{ij} = \left( \alpha_i - \alpha_j\right)$.
Approximating $\bar T_{ij} = T_{c,ij}$, eq.~(\ref{eq:seebeck_term})
simplifies to
\begin{equation}
  \alpha_{ij} = \bar\alpha_{ij} =\frac{\alpha_i + \alpha_j}{2}.
\label{eq:alpha}
\end{equation}

For the Peltier coefficient $\Pi_{ij}$, finally, one has to take the
Peltier heat properly into account, which is delivered to or taken from the
adjacent materials $i$ and $j$ at the interface. The electrical current $I_{ij}$
carries the heat $\Pi_iI_{ij}$ away from site $i$ and delivers
$\Pi_jI_{ij}$ to site $j$. We assume that the energy difference 
$(\Pi_i-\Pi_j)I_{ij}$, which is set free (or consumed) at the
interface, is given to (or taken from) both adjacent sites in equal
parts. The net heat current induced by $I_{ij}$ is then
\begin{align}
\Pi_{ij}I_{ij}&=\Pi_iI_{ij}-\frac{1}{2}\left(\Pi_i-\Pi_j\right)I_{ij}\\
&=\Pi_jI_{ij}+\frac{1}{2}\left(\Pi_i-\Pi_j\right)I_{ij}
\end{align}
with 
\begin{equation}
\Pi_{ij}=\frac{\Pi_i+\Pi_j}{2}.
\end{equation}
Due to the Kelvin relation, $\Pi_i=T_i\alpha_i$ for material $i$.

For the temporal evolution of the local temperatures, which is
essential for the Harman method, the heat capacities $C_i$ must be
given for the material belonging to site $i$.  Without specifying the
material further, the ratio $C/K$ is the typical time scale, on
which temperature differences between neighboring sites are leveled
out in the network by heat conduction. On the other hand, the
electrical site capacitances $C_{\text{el}}$ can be neglected, because
the equilibration time $C_{\text{el}}/G$ of the electrochemical potential
is much shorter than $C/K$. Therefore it is a reasonable
approximation that the electrostatic potential adjusts instantaneously
compared to the slow thermal evolution of the system.

The time evolution of the temperature $T_i$ is described by
\begin{equation}
  \dot T_i = \frac{1}{C_i} \sum_j\left(-I_{q,ij} 
    + I_{ij}\frac{\left(\mu_i - \mu_j\right)}{2 q}\right)\label{eq:dgl_temp}.
\end{equation}
The second term is the Joule heat (eq.~(\ref{eq:div_heat})) set free
on bond $ij$. The factor 1/2 accounts for its delivery in equal parts
to sites $i$ and $j$. For the numerical integration the time step is
about one-hundredth of the time scale of the fastest heat exchanging
mechanism $C/K$.

\begin{figure}[]
  \includegraphics[width=0.45\textwidth]{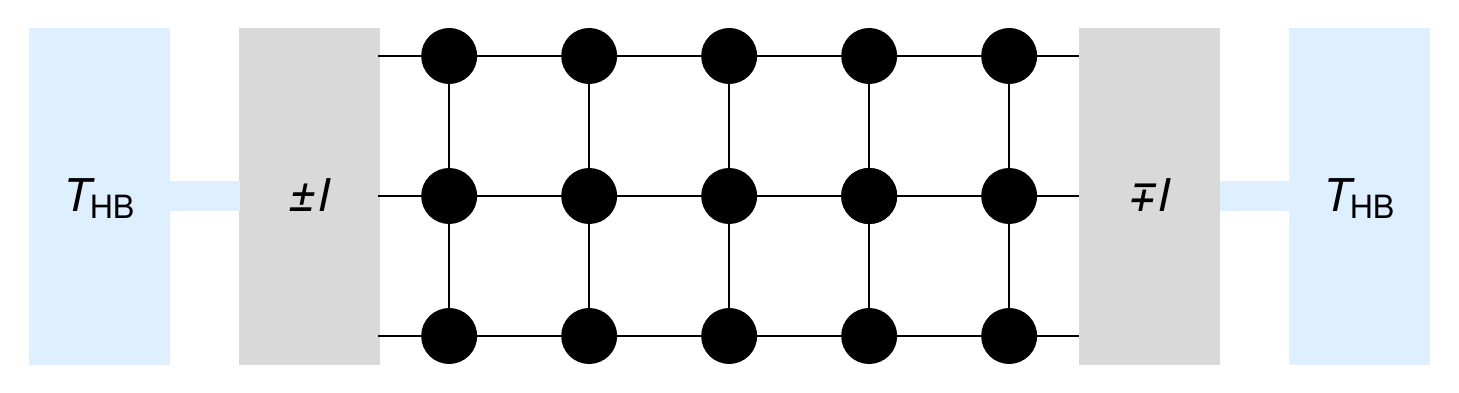} 
  \caption{(Color online) Two-dimensional sketch of
    setup H. It is composed of the material grid, the electrodes,
    where electric current $I$ is injected or extracted, and a heat
    bath (HB) at ambient temperature $T_{\text{HB}}$, coupled to the
    electrodes via heat conductances $K_{\text{HB}}$.}
  \label{fig:setup}
\end{figure}

The simulation procedure works as follows: An initial temperature
distribution and a total current $I$ entering one electrode and
leaving the other are provided (see Fig.~\ref{fig:setup}). The electrochemical
potentials are calculated from Kirchhoff's first law and then fed into
eq.~(\ref{eq:dgl_temp}) in order to calculate the temperatures for the next
time step. The temporal evolution of the temperatures implies
continuous adjustments of the local currents (according to
Eq.~(\ref{eq:el_curr_model})) and hence the
electrochemical potentials. Iterating these steps gives the transients
and finally the stationary state.

Electrodes are attached to the opposite surfaces normal to the
$x$-axis. They are assumed to have uniform electrochemical potential
and temperature. In this paper the electrical current $I$ flowing
through the material (and the electrodes) is fixed. Heat losses
through the surfaces parallel to the overall current direction are not
taken into account.  Two different boundary conditions for the
temperature at the electrodes are considered. In Sections
\ref{sec:cpm} and \ref{sec:segmented} the temperatures
$T_0>T_L=300\,\mr{K}$ of the electrodes at $x=0$ and $x=L$ are
fixed. We denote this as ``setup C''. In the Sections starting with
\ref{sec:Harman_method}, a different ``setup H'' is used (see
Fig.~\ref{fig:setup}): The electrodes are in contact with the
environment, which acts as a heat bath of fixed temperature
$T_{\mathrm{HB}}=300\,\mr{K}$. The temperatures of the electrodes are
determined by Joule heating, Peltier effect and heat exchange with the
environment. In general they differ from $T_{\mathrm{HB}}$ and depend
on the current direction.

\section{Application}
\subsection{Constant property model (CPM)}
\label{sec:cpm}

Here we briefly recall the analytical solution of the one-dimensional
CPM \cite{Seifert2006,Goupil2011}.  Equations (\ref{eq:j}) -
(\ref{eq:div_heat}) yield in steady state
\begin{equation}
  \frac{\partial^2 T(x)}{\partial x^2} =  
  -\frac{j^2}{\sigma \kappa} \label{eq:temp_dgl},
\end{equation}
where $j$ represents the $x$-component of the electrical current density. For fixed
temperatures $T_0$ and $T_L$ the solution reads
\begin{equation}
  T(x)=-\frac{j^2}{2\sigma \kappa} x(x-L) + \frac{T_L-T_0}{L} x +T_0\label{eq:temp}.
\end{equation}
The symmetry of the $j^2$-term with respect to $x=L/2$ implies that
the Joule heat produced in the sample is flowing equally to both electrodes.
Similarly, the electrochemical potential profile $\mu(x)$ follows from
eq.~(\ref{eq:j}) and eq.~(\ref{eq:temp})
\begin{equation}
  \frac{\mu(x)}{q} = \frac{\alpha j^2}{2\sigma \kappa} x(x-L) -
  \left(\frac{j}{\sigma}+ \alpha \frac{T_L -T_0}{L} \right) x .
\label{eq:pot}
\end{equation}
Without loss of generality, the integration constant in (\ref{eq:pot})
is $\mu(0)=\mu_0=0$.

\begin{table}[]
  \begin{tabular}{c c c c c c}
    \hline
    \hline
    &$\alpha [\mathrm{V/K}]$ & $\sigma [\mathrm{S/m}]$ & $\kappa [\mathrm{W/(K m)}]$\\ 
    \hline
    \rule{0pt}{2.5ex}   material A & $1 \cdot 10^{-4}$ &  $ 10^5$ & $2$ \\
    \hline
    \rule{0pt}{2.5ex}   material B & $2 \cdot 10^{-4}$ &  $10^4$ & $1$ \\

\end{tabular}
\caption{The parameters used for the simulations of segmented CPM
  thermoelectrics. Furthermore we set $T_{\mr{HB}}=300\,\mr{K}$ and $K_{\mr{env}}=0.01\,\mr{W/K}$.}
\label{tab:segmented_params}
\end{table}

\subsection{Segmented thermoelectrics}
\label{sec:segmented}
\begin{figure}[tbp]
  \includegraphics[width=0.45\textwidth]{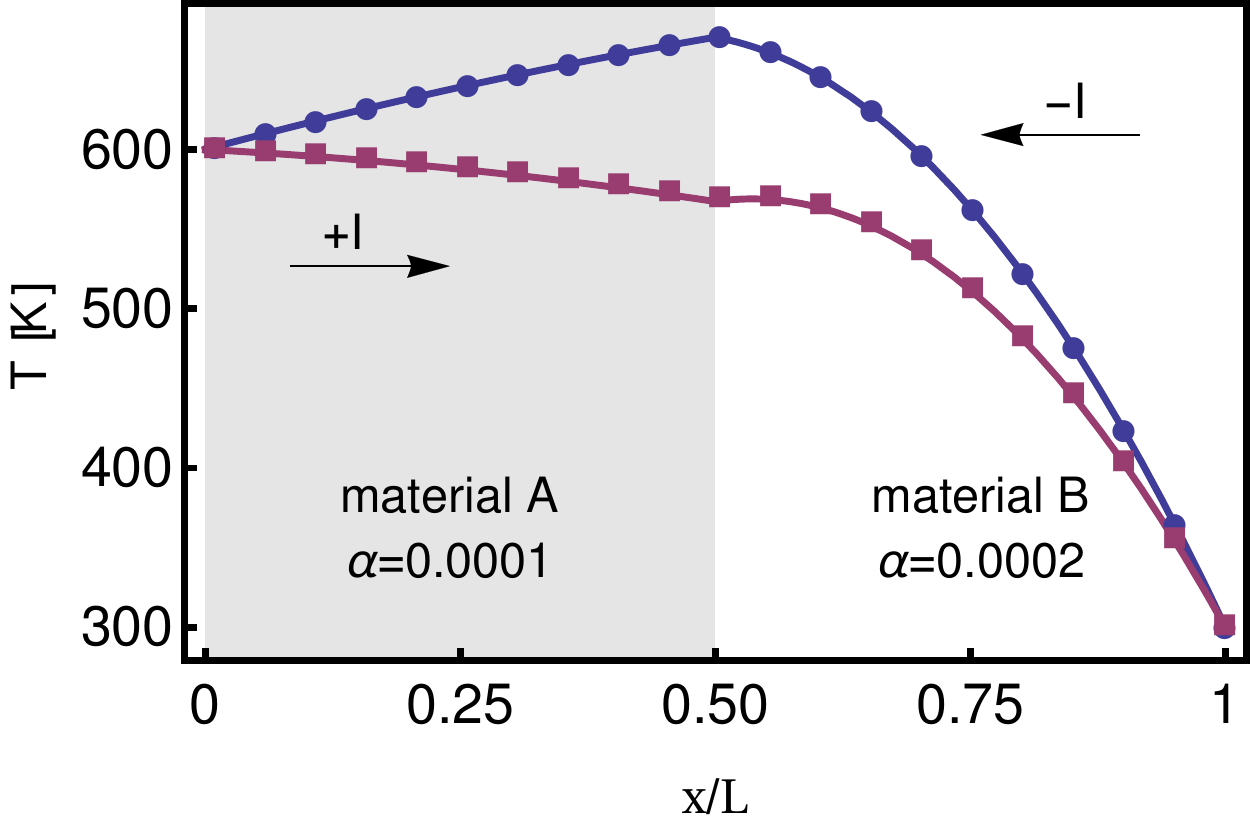} 
  \caption{(Color online) Temperature profile in a layered
    heterogeneous material calculated as in
    \cite{Yang2012,Mueller2006} (lines) and the simulation result
    (symbols) for $I=-50\,\mathrm{A}$ (points) and $I=50\,\mathrm{A}$
    (squares).}
  \label{fig:temp_prof_layered}         
\end{figure}
Segmented thermoelectrics consist of two layers of different
material. If represented by a one-dimensional piecewise constant
property model, analytic expressions for the
temperature and potential profile, as well as for the effective
transport coefficients can be obtained \cite{Mueller2006}. 
These expressions shall be compared to the corresponding simulation
results in order to validate the model presented in Sec.~\ref{sec:model}.

The analytical calculation makes use of the fact that for each segment
Eq.~(\ref{eq:temp_dgl}) holds, so that the temperature profile is
piecewise parabolic with the respective parameter sets.  The six
integration constants (for each material two from
Eq.(\ref{eq:temp_dgl}) and one from $\partial\mu/\partial x$) are
fixed by the boundary conditions $T_0$, $T_L$ and $\mu_0=0$ and by
requiring continuity of $T(x)$, $\mu(x)$ and $j_{{\text q},x}(x)$ at the
interface. 

Figure \ref{fig:temp_prof_layered} shows the temperature profile for
setup C with the parameters given in table
\ref{tab:segmented_params}. The electrodes are characterized by the
same parameters as the adjacent material, which prevents additional
Peltier heating/cooling at the electrode-sample interfaces. The size
of the sample is $L=L_x=L_y=L_z=10^{-2}\,\mathrm{m}$, discretized by
$N_y = N_z = 1$, $N_x =100$ lattice sites.  The current
$|I|=50\,\mathrm{A}$ flows from left to right (squares) or in reverse
direction (blue points). The effect of Peltier heating/cooling at the
interface can be clearly seen: The temperature profiles differ
significantly for the two current directions. It is this internal
temperature profile which affects the results of the Harman method, as
will be discussed in Sec.~\ref{sec:harman_segmented}.

Considering the bond at the interface where $\Delta \alpha_{ij} \ne
0$, we note that eq.~(\ref{eq:alpha}) is an approximationl, if a
non-linear temperature distributions is present. It modifies the total
voltage by an interface term, the relative contribution of which
vanishes if $N_x$ becomes large. As we discarded interface resistances
previously, this interface effect may be neglected as well.

\subsection{Harman method}
\label{sec:Harman_method}

\begin{figure}[]
  \includegraphics[width=0.5\textwidth]{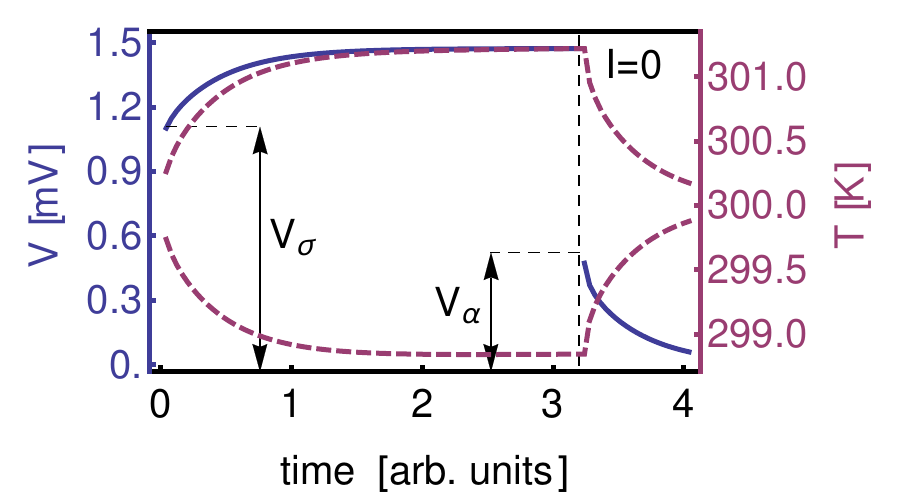} 
  \caption{(Color online) Simulated time-dependency of the voltage
    between the electrodes (left, solid line) and electrode
    temperature temperatures (right, dashed lines).}
  \label{fig:volt_time}         
\end{figure}

The Harman method \cite{Harman1958,Harman1959} is a measurement
technique, which allows to determine the three transport parameters
and consequently the figure of merit
\begin{equation}
  ZT = \frac{\alpha^2 \sigma}{\kappa}T = \frac{\alpha^2 G_{\text{tot}}}{K_{\text{tot}}}T
  \label{eq:zT}
\end{equation}
within one single measurement procedure. A setup H as shown in
fig.~\ref{fig:setup} is used, and a known dc current $I$ is
applied. Due to the Peltier effect one electrode/sample boundary heats
up and the other one cools down. At the same time, Joule heating leads
to an additional change of the sample temperature. This continues
until a steady state is reached, where heat conduction compensates
further temperature changes. The time evolution of the electrode
temperatures and the voltage across the sample (see
fig.~\ref{fig:volt_time}) are measured. After reaching the steady
state the current is switched off.

Starting with a homogeneous temperature the electric
conductance can be determined from a
measurement of the voltage $V_\sigma$ (see fig.~\ref{fig:volt_time})
via
\begin{equation}
  G_{\mathrm{tot}}' = \frac{I}{V_\sigma},
  \label{eq:G_tot}
\end{equation}
Primed quantities denote Harman method measurement results.  In
experiments this measurement might be difficult, as the Peltier effect
starts immediately creating a temperature difference between the
electrodes, when the current is switched on \cite{Harman1959}.
However, determining $G_{\mathrm{tot}}$ during simulations is simpler,
since $V_\sigma$ is recorded before the temperature change affects the
voltage. The Seebeck coefficient
\begin{align}
 \alpha' = \frac{V_\alpha}{\Delta T}.
 \label{eq:seebeck}
\end{align}
is calculated from the temperature difference $\Delta T$ and the
Seebeck voltage $V_\alpha$ measured immediately after switching off
the current $I$.

The heat conductance of the sample, 
\begin{equation}
  K_{\mathrm{tot}}' = \frac{\alpha' \bar{T} + (\bar{\mu} -
    \bar{\mu}_{\mr{env}})/q}{T_L-T_0} I - \frac{K_{\text{env}}}{2} \ ,
\label{eq:harman_heat_conductance}
\end{equation}
is obtained from energy balance in the steady state \cite{Harman1959}.
Here, $K_{\text{env}}$ denotes the heat conductance between the heat bath
and the electrodes. The bars indicate averaging of the values at
$x=0$ and $x=L$, e.g. $\bar{T}=(T_0 + T_L)/2$. 
 $\mu_{{\text{env}},0}$ and $\mu_{{\text{env}},L}$ are
the electrochemical potentials of the leads next to the left and right
electrode, respectively.

Let us briefly recall the derivation of
(\ref{eq:harman_heat_conductance}) according to \cite{Harman1959}.
The energy current $I_{\text e}=I_{\text q}+\mu/q I$ arriving at the left
electrode from the environment (heat bath or lead) is
\begin{equation}
  I_{\text{e,in}}(0) = \frac{\mu_{{\text{env}},0}}{q} I +K_{\text{env}}
  \left(T_{\text{HB}}-T_0\right)+\frac{I^2}{G_{\text{env}}}.
  \label{eq:energy_flow_wires_cold}
\end{equation}
In steady state it must be equal to the energy current 
\begin{equation}
  I_{\text e,out}(L) = \frac{\mu_{{\text{env}},L}}{q}I + K_{\text{env}}\left(T_L-T_{\text{HB}}\right)-\frac{I^2}{G_{\text{env}}},
  \label{eq:energy_flow_wires_hot}
\end{equation}
leaving the sample on the right hand side. In these expressions, the
last terms are the fractions of Joule heat produced in the leads,
that flow into the sample. They are assumed to be equal for
simplicity. The Peltier coefficient of the (metallic) leads is
assumed to be negligibly small. Taking the average of these
expressions eliminates the parameters $T_{\text{HB}}$ and $G_{\text{env}}$:
\begin{equation}
  \bar{I_{\text e}} = \frac{\bar{\mu}_{\text{env}}}{q} I + \frac{K_{\text{env}}}{2}\left(T_L - T_0\right).
\label{eq:average_energy_current_a}
\end{equation}
On the other hand, the energy current entering the sample from the
left must be equal to (\ref{eq:energy_flow_wires_cold}).  In the
framework of the constant property model it is given by
\begin{equation}
  I_{\text{e,out}}(0) = \left(\alpha T_0 + \frac{\mu_0}{q} \right) I + K_{\text{tot}}
  \left(T_0-T_L\right)-\frac{I^2 }{2 G_{\text{tot}}},
  \label{eq:energy_flow_cold}
\end{equation}
where the subscripts refer to the values at $x=0$ or $x=L$,
respectively, as before. This in turn must be equal to the energy current
arriving at the right electrode from the sample,
\begin{equation}
  I_{\text{e,in}}(L) = \left(\alpha T_L + \frac{\mu_L}{q} \right)I + K_{\text{tot}}
  \left(T_0-T_L\right)+\frac{I^2}{2 G_{\text{tot}}}.
  \label{eq:energy_flow_hot}
\end{equation}
Again taking the average of these two expressions eliminates the
parameter $G_{\text tot}$:
\begin{equation}
\bar{I_{\text e}} = \left(\alpha\bar{T}+\bar{\mu}/q\right)I - K_{\text{tot}}\left(
T_L - T_0\right).
\label{eq:average_energy_current_b}
\end{equation}
Identifying this with (\ref{eq:average_energy_current_a}) leads to
(\ref{eq:harman_heat_conductance}).  ($\bar{\mu} -
\bar{\mu}_{\text{env}})/q$ is the average contact potential between
the electrodes and the leads. Here we do not take contact resistances
into account. Hence it does not enter the evaluation of the simulation
data. In experiments additional losses due to convection and heat
radiation occur, which can be taken into account via correction terms
added to eq.~(\ref{eq:harman_heat_conductance}) (e.g. \cite{Ao2011}).
\begin{figure*}[]
  \subfloat[electric conductivity]{\label{fig:eff_seg_a}\includegraphics[width=0.45\textwidth]{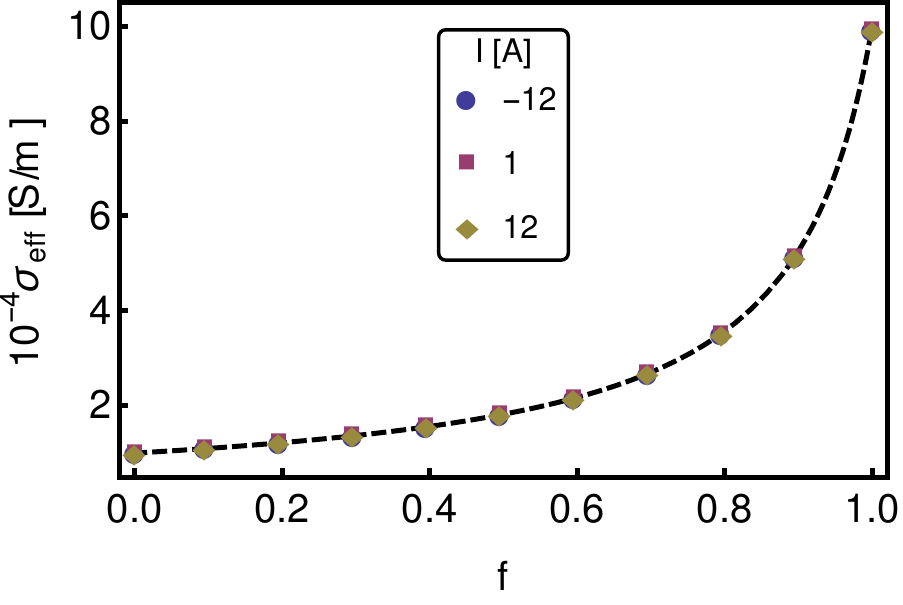}}
  \subfloat[Seebeck coefficient]{\label{fig:eff_seg_b}\includegraphics[width=0.45\textwidth]{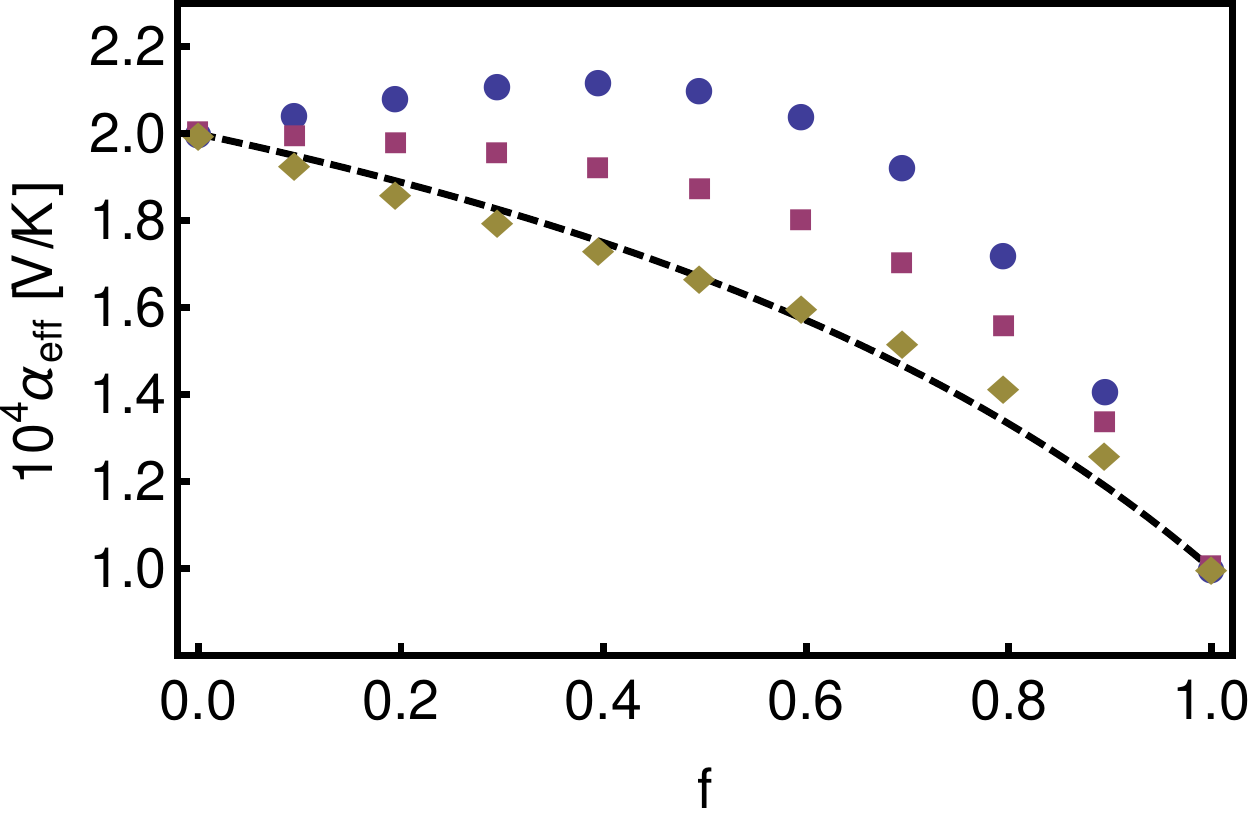}}\\
  \subfloat[heat conductivity]{\label{fig:eff_seg_c}\includegraphics[width=0.45\textwidth]{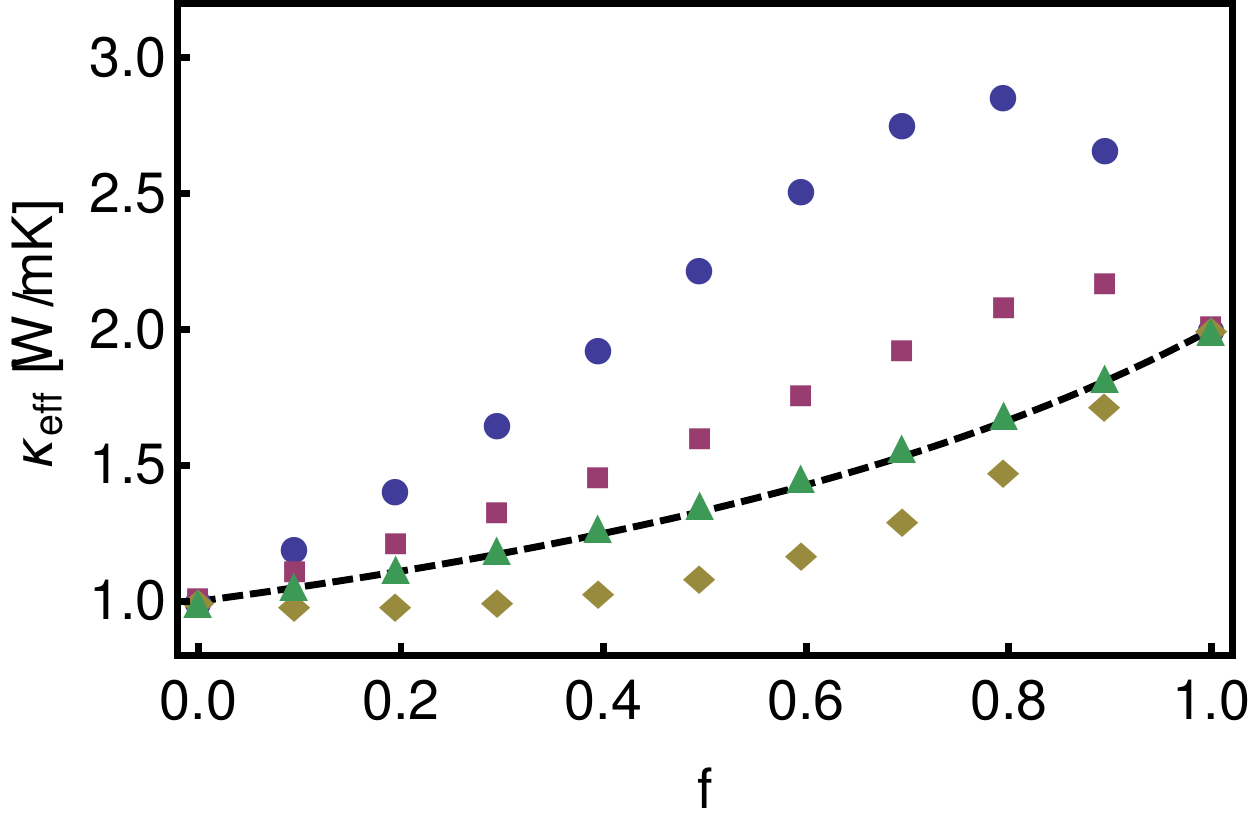}}
  \subfloat[figure of merit]{\label{fig:eff_seg_d}\includegraphics[width=0.45\textwidth]{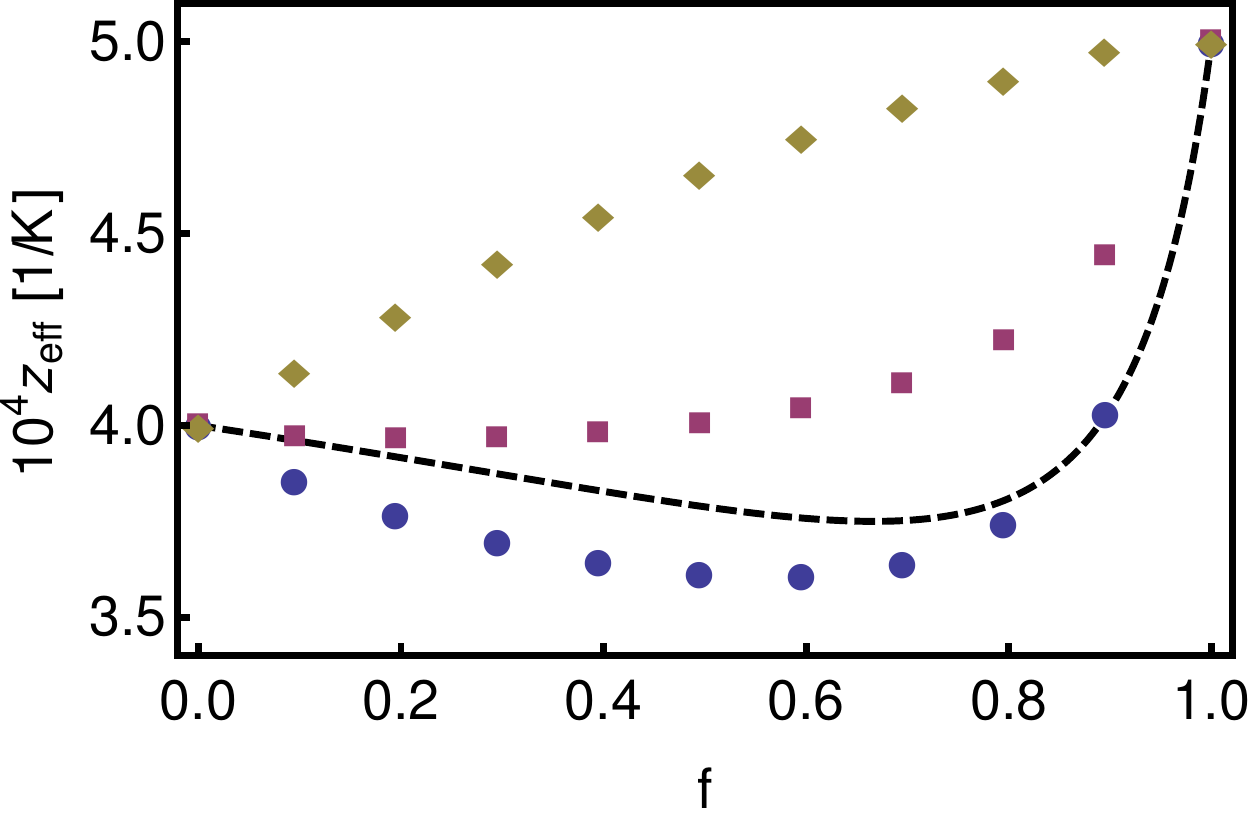}}
  \caption{(Color online) The effective transport coefficients and
    $z_\eff=(\alpha_{\text{eff}}^2\sigma_{\text{eff}})/\kappa_{\text{eff}}$
    in dependence of the volume fraction $f$ of material A in a
    segmented thermoelectric with the parameters of table
    \ref{tab:segmented_params}. The dashed lines represent the
    effective expressions eqs.~(\ref{eq:eff_elec_cond}),
    (\ref{eq:eff_heat_cond2}),(\ref{eq:eff_seebeck}), while the dots
    are obtained from simulations. The solid lines in the upper right
    figure are analytic results discussed in the text.}
  \label{fig:effective_parameters}         
\end{figure*}

The Harman method gives correct results for homogeneous
samples. However, as we are going to explain in the next two sections,
when applied to heterogeneous systems, it can give values for the
Seebeck coefficient, which can be more than 100\% off the true value.
Consequently, also the heat conductance derived from
(\ref{eq:harman_heat_conductance}) will be misleading.

\subsection{The Harman method applied to segmented thermoelectrics}
\label{sec:harman_segmented}

The basic problem encountered, when the Harman method is applied to
inhomogeneous systems, will be explained in this section for the
example of a bilayer thermoelectric. We simulated a setup H
(fig.~\ref{fig:setup}) consisting of materials A and B with parameters
given in table~\ref{tab:segmented_params}. The volume fraction of the
A-layer is denoted by $f$. The electrodes are characterized by the
same parameters as the adjacent layers, except of the Seebeck
coefficient, which is set to zero. The dimensions are chosen as in
sec.~\ref{sec:segmented}.

The transport coefficients are measured with the Harman method 
for different ratios $f$ and currents $I$. The results are then
compared to the analytical expressions, which will be given first. 

Denoting the electrochemical potential at the interface betweem the A-
and B-segment by $\mu_{\mr AB}$, the electrical conductivity of
material A in the absence of a temperature gradient is
\begin{equation}
  \sigma_{\text A} = \frac{q j\ fL}{\mu_0 - \mu_{\text{AB}}}.
  \label{eq:sigma_A}
\end{equation}
The heat conductivity, on the other hand, is defined under open
circuit conditions, $j=0$, as
\begin{equation}
  \kappa_{\text A}=\frac{j_q\ fL}{T_0-T_{\text{AB}}},
  \label{eq:kappa_A}
\end{equation}
with the temperature $T_{\text{AB}}$ at the interface. The expressions
for $\sigma_{\text B}$ and $\kappa_{\text B}$ read accordingly. The
effective electrical and heat conductivities are then derived as a
connection in series:
\begin{eqnarray}
  \sigma_{\eff} &=& \frac{q j\ L}{\mu_0 - \mu_{\text L}}= \frac{\sigma_{\text
      A} \sigma_{\text B}}{\sigma_{\text A}(1-f)+\sigma_{\text B} f},
  \label{eq:eff_elec_cond}\\
  \kappa_{\eff} &=& \frac{j_q\ L}{T_0-T_{\text L}} = \frac{\kappa_{\text
      A} \kappa_{\text B}}{\kappa_{\text A} (1-f) + \kappa_{\text B} f}.
  \label{eq:eff_heat_cond2}
\end{eqnarray}

The effective Seebeck coefficient, also defined for open circuit
conditions, corresponds to a series connection of the Seebeck voltages
created from material A and B:
\begin{eqnarray}
  \alpha_{\eff} &=& -\frac{1}{q}\left(\frac{\mu_0 - \mu_L}{T_0 -
      T_L}\right) \nonumber \\ 
  &=& \frac{\kappa_A(1-f) \alpha_B + \kappa_B f \alpha_A}{\kappa_A
    (1-f)+\kappa_B f}.
  \label{eq:eff_seebeck}
\end{eqnarray}
\begin{figure}[]
  \centering
  \includegraphics[width=0.45\textwidth]{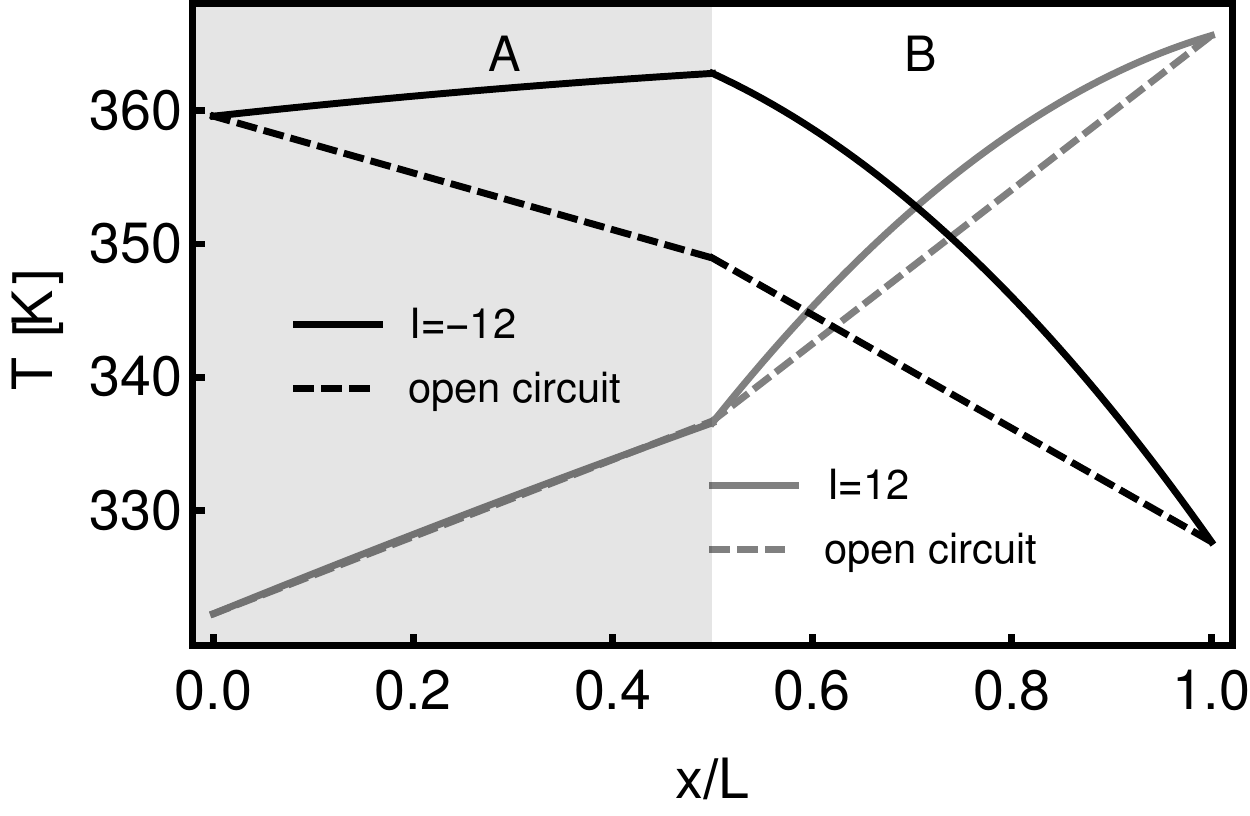}
  \caption[Temperature distribution]{The figure shows the temperature
    distribution for $f=0.5$, $I=12\,\mathrm{A}$ (gray, solid line)
    and $I=-12\,\mathrm{A}$ (black, solid line). Additionally, the
    temperature distribution in open circuit conditions are plotted
    using the same boundary temperatures (dashed lines).}
  \label{fig:harman_temp_dist}         
\end{figure}
Fig.~\ref{fig:effective_parameters} shows a comparison between
eqs.~(\ref{eq:eff_elec_cond}), (\ref{eq:eff_heat_cond2}),
(\ref{eq:eff_seebeck}) and the results of a simulation of the Harman
method. As expected, the electric conductivity obtained from the
Harman method coincides perfectly with the analytic results
(fig.~\ref{fig:effective_parameters}, upper left). But the Seebeck
coefficient (upper right) and the heat conductivity measured via the
Harman method depend on the current, which was applied before the
measurement of $V_\alpha$ and deviate strongly from the open circuit
values, eqs.~(\ref{eq:eff_heat_cond2}) and (\ref{eq:eff_seebeck}).
Only in the limiting cases $f\rightarrow 1$ and $f\rightarrow 0$ the
differences vanish. Consequently, the same is true for $z_\eff$.

Considering the Seebeck coefficient, the deviations arise from the
influence of the Peltier heating/cooling at the interfaces on the
temperature profile (see fig.~\ref{fig:harman_temp_dist}).  It causes
the current dependence of the temperatures $T_0,\ T_L$, $T_{\mr{AB}}$
and hence of the thermopower
\begin{equation}
  \label{eq:seebeck_voltage}
  V_\alpha = \alpha_A \left(T_{\mr{AB}} -T_0 \right) + \alpha_B \left(T_L-T_{\mr{AB}}\right).
\end{equation}
If one compares the analytic temperature profiles for $f=1/2$ created
by the external current $I=-12\,\mathrm{A}$ (black, solid line) and
$I=12\,\mathrm{A}$ (gray, solid line) to the respective open circuit
profiles (dashed lines) with the same boundary temperatures, a much
better agreement, particularly of the interface temperature
$T_{\mr{AB}}$, is found for $I=12\,\mathrm{A}$, because Joule heating
is nearly compensated by Peltier cooling.  This explains, why the
effective Seebeck coefficients for $I=12\,\mathrm{A}$ almost coincide
with eq.~(\ref{eq:eff_seebeck}), while for $I=-12\,\mathrm{A}$ the
Seebeck coefficient is strongly overestimated, since the larger
temperature drop is across the material with higher Seebeck
coefficient.

\begin{figure}[]
  \centering
  \includegraphics[width=0.45\textwidth]{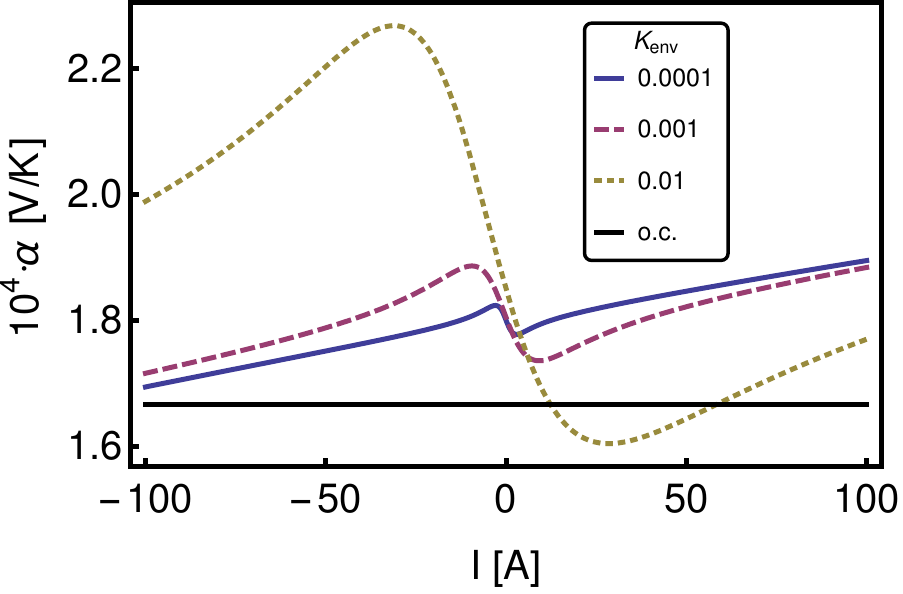}
  \caption{(Color online) Seebeck coefficient measured with the Harman
    method for varying currents $I$ and heat coupling $K_{\mr{env}}$
    of a segmented thermoelectric with $f=0.5$. The black line
    corresponds to the correct open circuit Seebeck coefficient. As
    the current approaches $I=0$ the difference between Harman and
    open circuit methods does not vanish.}
  \label{fig:seebeck_varI}         
\end{figure} 

Although the Seebeck coefficient is measured at $I=0$ when using the
Harman method, it depends strongly on the previous current (see
fig.~\ref{fig:seebeck_varI}), because the temperature profile in
inhomogeneous samples acts as a memory. Remarkably, even for
arbitrarily small currents $|I|$ the Harman measurement gives a
Seebeck coefficient which deviates from the true open circuit
value. In this limit Joule heating is negligible and the temperature
at all three interfaces depend linearly on the current due to Peltier
interface effect. Hence, the relation between the temperature
differences across both materials becomes independent of $I$ for $I\to
0$.

\begin{figure}[]
  \centering
  \includegraphics[width=0.48\textwidth]{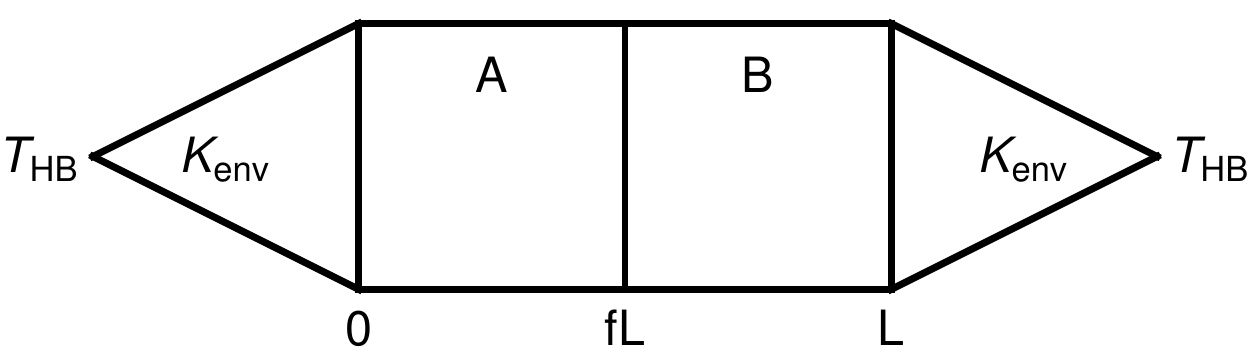}
  \caption{An segmented structure connected to heat bathes of
    temperature $T_{\mr{HB}}$ via the heat conductance
    $K_{\mr{env}}$.}
  \label{fig:segmented_analytic_setup}         
\end{figure}

We elaborate this argument by deriving the Seebeck coefficient of a
segmented thermoelectric measured by the Harman method
analytically. For this purpose we consider the AB structure depicted
in fig.~\ref{fig:segmented_analytic_setup}, which is connected to heat
bathes with fixed temperature $T_{\mr{HB}}$ via heat conductance
$K_{\mr{env}}$. For each segment $i$ the temperature follows
\begin{equation}
  T_i(x)= a_i x^2+ b_i x +c_i \quad \mathrm{with} \quad a_i= -\frac{j^2}{2\sigma_i\kappa_i}
\end{equation}
Finding the unknown $b_i,c_i$ works basically as discussed in
sec.~\ref{sec:segmented}, whereby we use a heat flux as boundary
conditions here.  Note that this procedure can easily be extended to
larger number of segments.  From $T(x)$ we derive the Seebeck
coefficient according to eq.~(\ref{eq:seebeck}) using
eq.~(\ref{eq:seebeck_voltage}), which coincides with the simulations
(lines in upper right of
fig.~\ref{fig:effective_parameters}). Usually, small currents are
applied allowing for a expansion of $T(x)$ around $j=0$, which leads
to
\begin{align}
  \label{eq:seebeck_harman_analytic}
&  \alpha_\eff'-\alpha_\eff = \nonumber\\ &\frac{
    (\alpha_A-\alpha_B)^2 (1-f)f\left(\kappa_\eff + K_{\mr{env}} L/(2S)
    \right)}{\alpha_\eff ((1-f)\kappa_A + f \kappa_B)} +\mathcal{O}(j)
\end{align}
with the cross section area of the sample $S$.  It is important to
note that the difference always has the same sign as $ \alpha_\eff$
meaning that the Harman method systematically overestimates the
absolute value of the Seebeck coefficient.

Measuring the heat conductivity according to
eq.~(\ref{eq:harman_heat_conductance}) leads to flawed results in a
segmented thermoelectric. 
However, we present a possibility to infer the heat conductivity
eq.~(\ref{eq:eff_heat_cond2}) from the Harman method requiring
$\alpha_\eff$, which can be gained from Harman measurments using
eq.~(\ref{eq:seebeck_harman_analytic}).  Basically, we repeat the
derivation for $K_{\mr{tot}}'$ as presented in
sec.~\ref{sec:Harman_method}.  We determine the energy currents at
$x=0$ and $x=L$ using the temperature distribution and its derivative
as derived for the strucure depicted in
fig.~\ref{fig:segmented_analytic_setup}.  The resulting energy
currents are expanded to first order around $j=0$ and their average
reads
\begin{align}
  \label{eq:harman_segmented_heat_cond}
   \bar I_e = \frac{K_{\mr{env}} \alpha_\eff T_0 j S }{2 K_{\mr{tot}} +
    K_{\mr{env}}} +\frac{\mu_0 + \mu_L}{2q} j S +\mathcal{O}(j^2).
\end{align}
Following the arguments of sec.~\ref{sec:Harman_method} this must be
equal to eq.~(\ref{eq:average_energy_current_a}), which leads to
\begin{align}
  \label{eq:heat_cond_harman_analytic}
  \kappa_{\eff} = \frac{\alpha_\eff T_{\mr{HB}} L}{(T_L-T_0)} j
  -\frac{K_{\mr{env}}L}{2S}+\mathcal{O}(j),
\end{align}
where we have set $\mu_0- \mu_{\mr{env},0} =\mu_L- \mu_{\mr{env},L} =
0$. Using $T_0$ and $T_L$ explicitly in
eq.~(\ref{eq:average_energy_current_a}), but not in
eq.~(\ref{eq:harman_segmented_heat_cond}) is not inconsistent at
all. In fact, we avoid products of the form $(T_0-T_{\mr{HB}}) j$ or
$(T_L-T_{\mr{HB}}) j$, which are of second order in $j$, since
$(T_0-T_{\mr{HB}})$ and $(T_L-T_{\mr{HB}})$ scale with $j$. The
application of eq.~(\ref{eq:heat_cond_harman_analytic}) leads to
satisfying agreement as shown by the green triangles in
fig.~\ref{fig:effective_parameters} (lower left) for $I=1\,\mr{A}$.

Finally, we discuss $z_\eff'$ determined by the Harman technique in a
segmented thermoelectric for $j\to 0$. Applying
eqs.~(\ref{eq:eff_elec_cond}),(\ref{eq:seebeck_harman_analytic}) and
determining the heat conductivity from
eq.~(\ref{eq:harman_heat_conductance}) with $\alpha_\eff'$, results in
\begin{align}
  \label{eq:z_harman_analytic_complete}
   z_\eff'-z_\eff = \frac{(\alpha_\eff +\alpha_c)^2-\alpha_\eff^2
    \frac{\kappa_\eff+\kappa_c}{\kappa_\eff}}{\kappa_\eff+\kappa_c}\sigma_\eff+\mathcal{O}(j)
\end{align}
with $\alpha_c=\alpha_\eff'-\alpha_\eff$ (see
eq.~(\ref{eq:seebeck_harman_analytic}) and $\kappa_c =
\alpha_cT_{\mr{HB}}j/(T_L-T_0)$.  According to
eq.~(\ref{eq:z_harman_analytic_complete}) the figure of merit may be
over or underestimated by the Harman method. Just for
$K_{\mr{env}}\to0$ we yield
\begin{align}
  \label{eq:z_harman_analytic}
  z_\eff'-z_\eff= \frac{(1-f)f \sigma_\eff(\alpha_A
    -\alpha_B)^2}{(1-f)\kappa_A - f \kappa_B}+\mathcal{O}(j) >0
\end{align}
Hence, for small currents and weak heat coupling the figure of merit
is always overestimated by the Harman technique.  Considering the
relative deviation $\Delta_z=(z'_\eff-z_\eff)/z_\eff$ in the limit
$K_{\mr{env}} \to 0$ we find that its maximum occuring at the most
unfavorable $f=f_{\mathrm{max}}$ is
\begin{align}
  \Delta_z(f_{\mathrm{max}}) = \frac{(\alpha_A-\alpha_B)^2}{4\alpha_A
    \alpha_B},
\end{align}
which is solely affected by the difference of both Seebeck
coefficients and can be very large. E.g. for
$\alpha_A=0.0004\,\mr{V/K}$ and $\alpha_B=0.0001\,\mr{V/K}$ we get
$\Delta_{z}(f_{\mathrm{max}}) = 0.5625$ and the Harman method
overestimates the $z$ by $56.25\%$.

We conclude that applying the Harman method to segmented structures as
done in \cite{Satake2004} is tricky and may lead to results that are
too optimistic.

\begin{figure}[]
  \centering
  \includegraphics[width=0.49\textwidth]{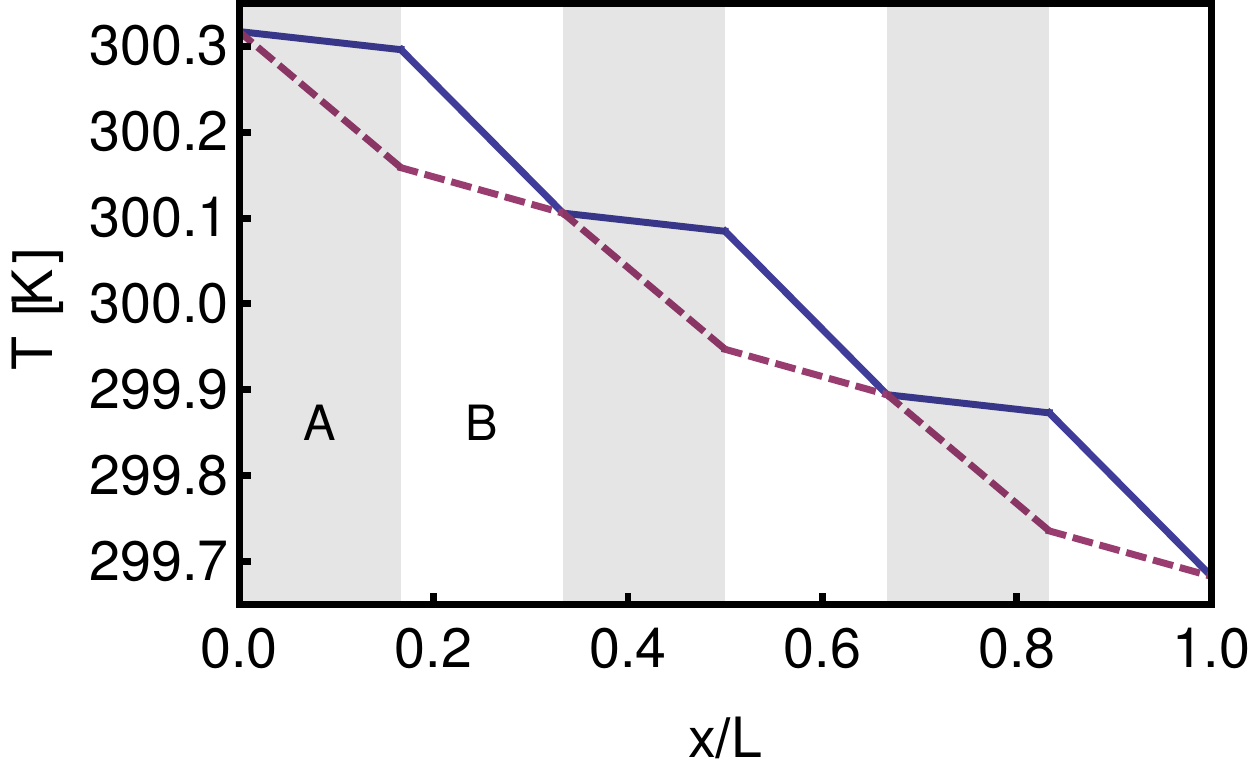}
  \caption{(Color online) First order temperature distribution (solid
    line) of a material consisting of $n=6$ segments and the open
    circuit temperature distribution (dashed line) with the same
    boundary temperatures. We used parameters from table
    \ref{tab:segmented_params}, except of $\kappa_A=1\,\mr{W/(K m)}$,
    $\kappa_B=3\,\mr{W/(K m)}$ and $K_{\mr{env}}=0.1\,\mr{W/K}$, in
    order to emphasize the effect discussed in the text.}
  \label{fig:temp_dist_six_segs}         
\end{figure} 
\begin{figure*}[]
  \subfloat[Seebeck
  coefficient]{\label{fig:eff_comp_a}\includegraphics[width=0.45\textwidth]{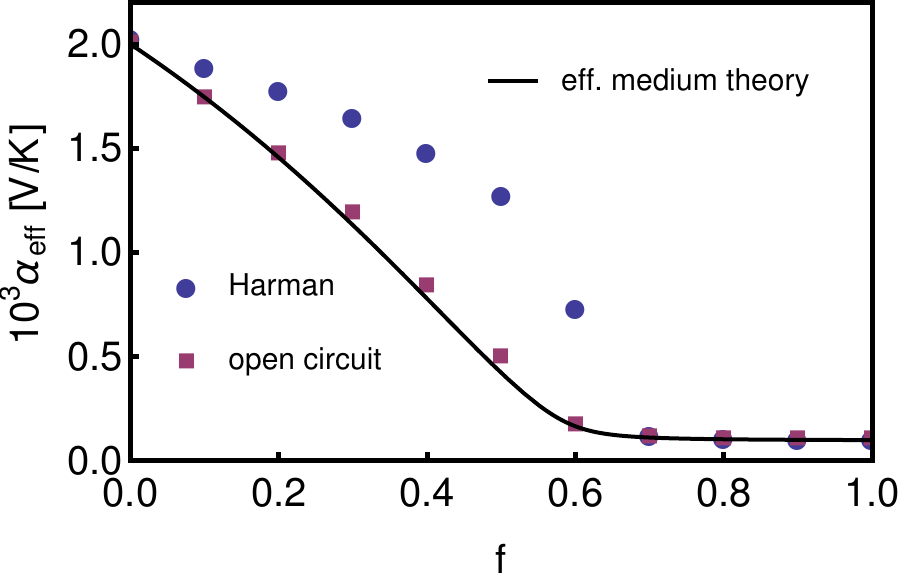}}
  \subfloat[heat
  conductivity]{\label{fig:eff_comp_b}\includegraphics[width=0.437\textwidth]{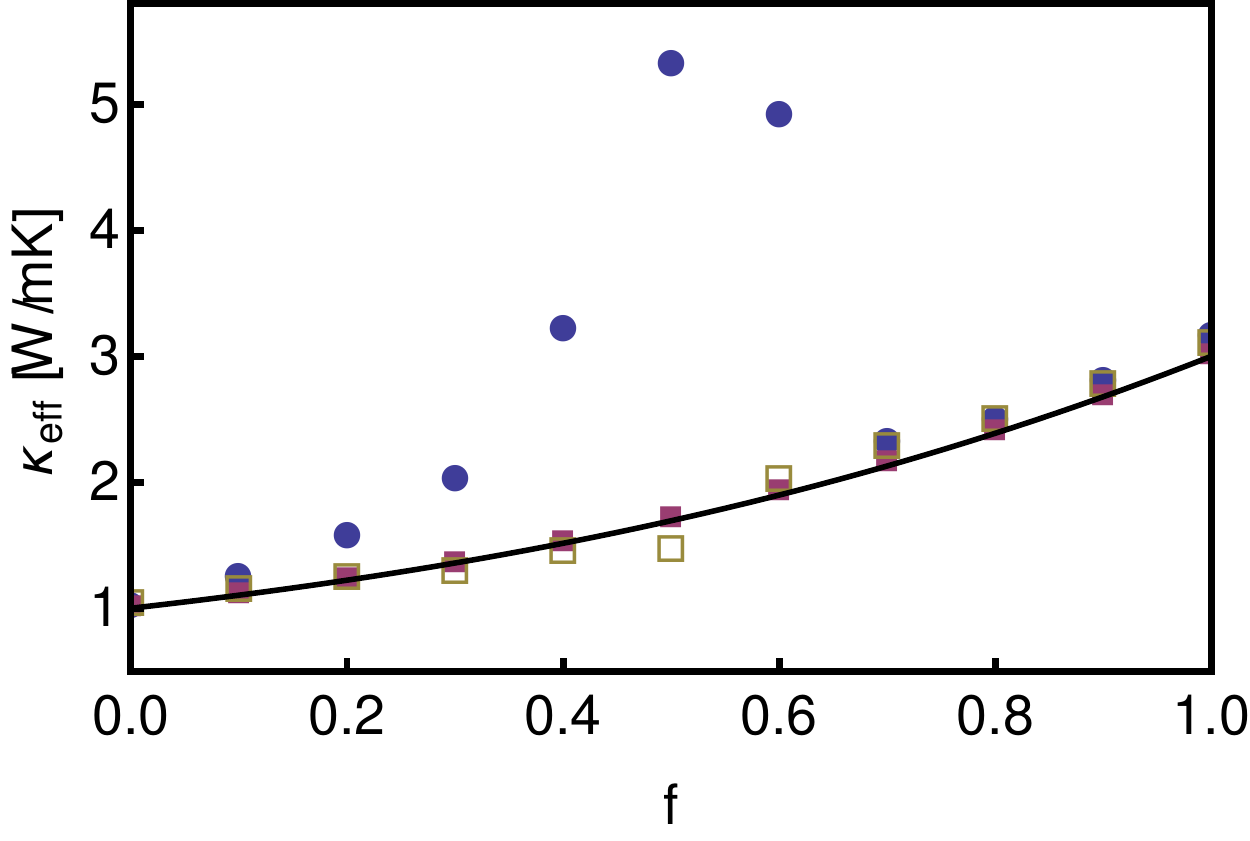}}
  \caption{(Color online) The Seebeck coefficient and the heat
    conductivity in dependency of the fraction $f$ of material A. The
    solid line always represents the corresponding analytic
    expression, while the dots are simulation results obtained by
    Harman method (dots) and open circuit measurements (squares). Open
    squares in fig.~\ref{fig:eff_comp_b} are Harman results using
    $\alpha_\eff$ from eq.~(\ref{eq:seebeck_comp}).}
  \label{fig:effective_cond_comp}         
\end{figure*}
Now we show that the systematic error, which we found, if the Harman
method is applied to a double-layer system, remains unchanged for
periodic superlattices in the limit of weak current. This is
particularly easy to see for even numbers of layers. First, for open
circuit conditions, a given temperature difference $T_L-T_0$ will be
evenly distributed among all $N$ double layers,
$T_{\nu+1}-T_{\nu}=(T_L-T_0)/N$ (see
fig.~\ref{fig:temp_dist_six_segs}). In first order this is also the
case, if a current is imposed. In particular, the temperature
dependence of the Peltier effect may then be neglected. Hence, each
double layer has the same systematic error of the effective Seebeck
coefficient, eq.~(\ref{eq:seebeck_harman_analytic}). Consequently the
homogeneous array of double layers gives rise to the same error.  If
the number of layers is odd, this conclusion remains true up to a
correction proportional to $1/N$ which vanishes for large
superlattices.

Superlattices attracted a lot of interest and a record $zT\approx2.4$
was measured using the Harman method in
$\text{Bi}_2\text{Te}_3/\text{Sb}_2\text{Te}_3$ superlattices
\cite{Venkatasubramanian2001}, which has not been reproduced so far.
The individual segments are very thin: $\mr{Bi}_2\mr{Te}_3$ segments
have a thickness of $1\,\mr{nm}$ and the $\mr{Sb}_2\mr{Te}_3$ segments
have a thickness of $5\,\mr{nm}$ implying $f=1/6$. Neglecting quantum
effects and using $\alpha_{\mr{BiTe}} \approx 2.2\cdot
10^{-4}\,\mr{V/K}$ \cite{Poudel2008}, $\alpha_{\mr{SbTe}} \approx
0.9\cdot 10^{-4}\,\mr{V/K}$ \cite{Mehta2012},
$\kappa_{\mr{BiTe}}\approx 2\,\mr{W/(m K)}$ \cite{Tritt2006} and
$\kappa_{\mr{SbTe}}\approx 1.8\,\mr{W/(m K)}$ \cite{Tritt2006} results
in $\Delta_{z}(1/6)\approx 0.18$ corresponding to $18\%$
overestimation of the figure of merit.

\subsection{Harman method and composite materials}

\begin{table}[]
  \begin{tabular}{c|c c c}
    & $\sigma$ [S/m] & $\kappa$ [W/(Km)] & $\alpha$ [V/T]\\[0ex]
    \hline
    \hline
      \rule{0pt}{2.5ex}  material A & $1.3\cdot 10^5$ & $3$& $ 0.0001$\\
      \rule{0pt}{2.5ex}  material B & $100$ & $1$& $ 0.002$\\
    \end{tabular}
    \caption{The parameters used for the simulation represent a metal-like (A) and a  semiconductor-like (B) material. Furthermore we set $T_{\mr{HB}}=300\,\mr{K}$ and  $K_{\mr{env}} = 0.01\,\mr{W/K}$. The system dimensions are as before $L=L_x=L_y=L_z=0.01\,\mr{m}$.}
  \label{tab:parameter_composite}
\end{table}

In this section, random composites made of domains of two different
materials are simulated using the network model. The effective
transport coefficients are determined from the Harman method, and
compared to the ones obtained under open circuit conditions and from
an effective medium theory.

Two-dimensional systems are considered with sizes up to
$40\times40$ sites, which are occupied randomly by two different types
of domains with metal-like and semiconductor-like transport
parameters given in tab.~\ref{tab:parameter_composite}. The fraction $f$
of the metal-like material (A) is varied and for each $f$ at least
$10$ different randomly chosen configurations were simulated and analyzed.
\begin{figure*}[]
 \includegraphics[width=0.05\textwidth]{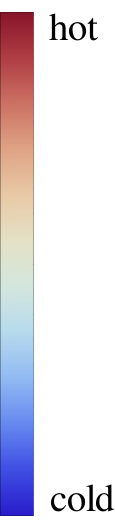}
  \subfloat[$f=0.8$]{\label{fig:dist_comp_a}\includegraphics[width=0.3\textwidth]{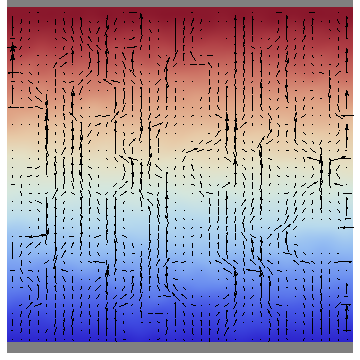}}
  \subfloat[$f=0.5$]{\label{fig:dist_comp_b}\includegraphics[width=0.3\textwidth]{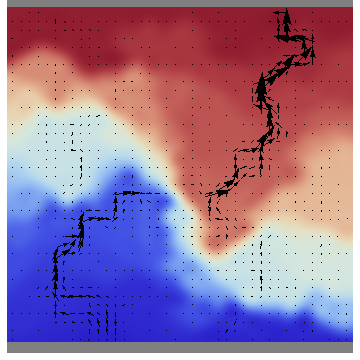}}
  \subfloat[$f=0.5$]{\label{fig:dist_comp_c}\includegraphics[width=0.3\textwidth]{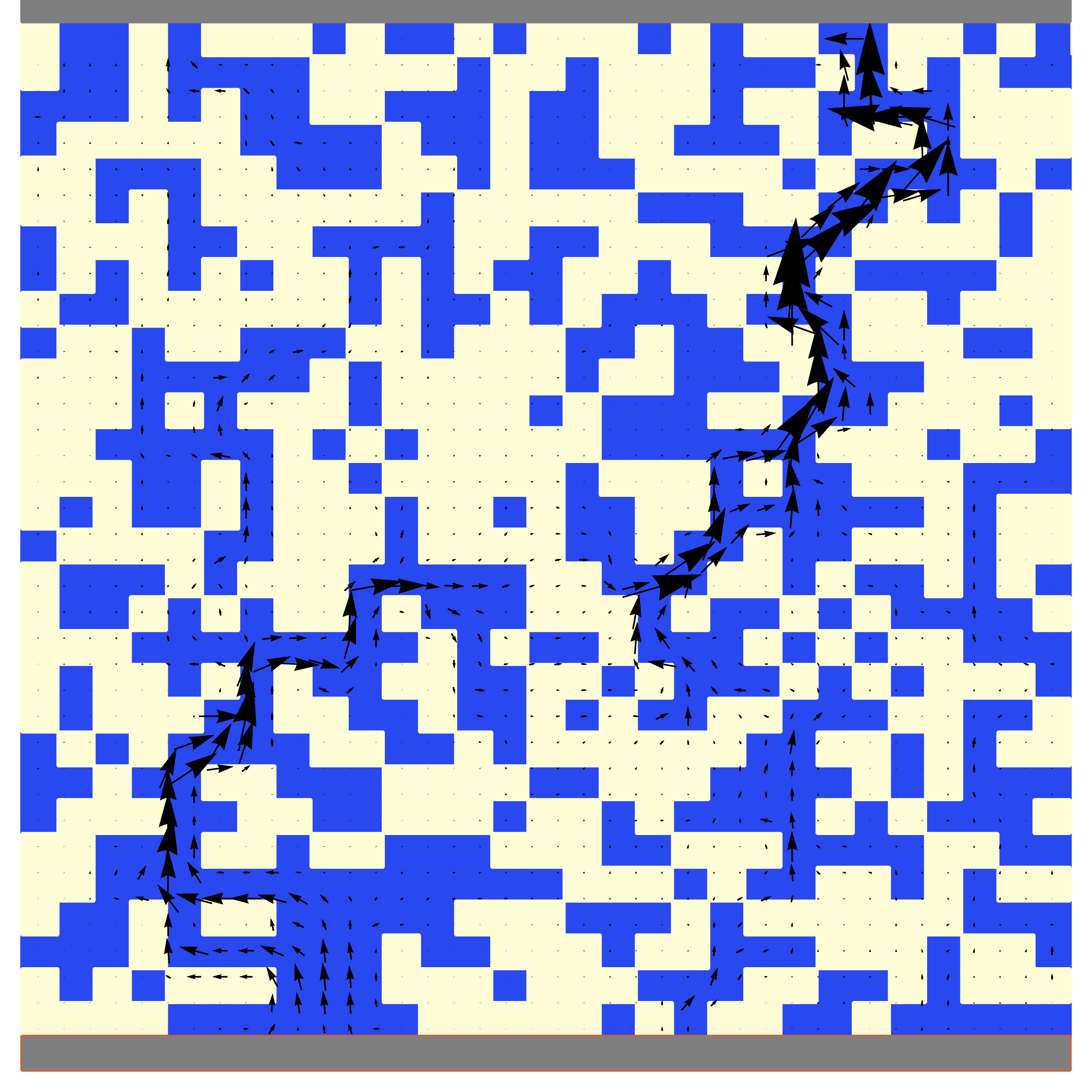}}
  \caption{(Color online) Fig.~\ref{fig:dist_comp_a} and
    fig.~\ref{fig:dist_comp_b} show temperature and current
    distribution close to ($f=0.5$) and far from ($f=0.8$) the
    percolation threshold for an example setup consisting of
    $30\times30$ lattice sites in a Harman setup at the moment of the
    Seebeck measurement.  The electrodes are printed at the bottom and
    the top of the samples, while the black arrows indicate the
    strength and the direction of the
    current. Fig.~\ref{fig:dist_comp_c} shows the particle
    distribution and blue (dark) squares represent metal-like and
    yellow (bright) squares represent semi conductor-like material.}
  \label{fig:dist_comp}         
\end{figure*}

In contrast to a segmented thermoelectric, the number of interfaces
between different materials is large and the error produced by the
approximation eq.~(\ref{eq:alpha}) might become relevant. We checked
that this is not the case by comparing with a simulation, in which
the Seebeck coefficient $\alpha_{ij}$ for a bond between the two
materials was determined from $\alpha_i$ and $\alpha_j$ weighted with
$K_i$ and $K_j$ as in eq.~(\ref{eq:eff_seebeck}). No significant
differences could be detected. Therefore we keep
eq.~(\ref{eq:alpha}) for the following simulations.

For such disordered two-component systems, an effective medium theory
was developed \cite{Mclachlan1987,Bergman1991,Bergman1999} with the
following expression for the Seebeck coefficient:
\begin{equation}
  \label{eq:seebeck_comp}
  \alpha_{\eff} = \alpha_B +\left(\alpha_A -\alpha_B\right)\frac{\kappa_\eff/\sigma_\eff - \kappa_B/\sigma_B}{\kappa_A/\sigma_A-\kappa_B/\sigma_B}.
\end{equation}
The effective medium electrical and heat conductivities are given by
\begin{equation}
\label{eq:eff_med}
(1-f)\frac{\sigma_B^{1/t}-\sigma_\eff^{1/t}}{\sigma_B^{1/t}+A \sigma_\eff^{1/t}}+f\frac{\sigma_A^{1/t}-\sigma_\eff^{1/t}}{\sigma_A^{1/t}+A\sigma_\eff^{1/t}}=0
\end{equation}
and an analogous equation for $\kappa_{\text{eff}}$. The parameter
$A=(1-f_c)/f_c$ is connected to the percolation threshold $f_c$. 

The Harman method gives an accurate effective electrical conductivity
for the composite material, as explained before. By fitting these
simulation results with eq.~\ref{eq:eff_med}, we determine the
effective medium parameters $f_c=0.594(2)$ and $t=1.315(8)$. Note that
$f_c$ is close to the expected percolation threshold $0.592746$ for
site percolation on a square lattice.  The effective medium theory
expressions for $\sigma_\eff$ and $\kappa_\eff$ with the above fitting
parameters are fed into eq.~(\ref{eq:seebeck_comp}) to obtain the
effective medium value for the Seebeck coefficient.

The Seebeck coefficient $\alpha_\eff'$ obtained by the Harman method
deviates strongly from the results for open circuit conditions and
from the effective medium theory, especially slightly below the
percolation threshold (see fig.~\ref{fig:effective_cond_comp}).  This
phenomenon can be understood by looking at the temperature and current
distributions at the moment the Seebeck coefficient is measured. For
$f = 0.8$, far above the percolation threshold for the metal-like
material, the temperature and current distribution appear homogeneous
(see fig.~\ref{fig:dist_comp_a}) leading to a good match of the Harman
and the open circuit measurements. In another sample at $f=0.5$ (see
fig.~\ref{fig:dist_comp_b} and \ref{fig:dist_comp_c}), however, a path
of well conducting material A almost percolates and thus carries a
majority of the current.  The percolation is interrupted by material B
and a strong Peltier heating/cooling appears. This together with
differences in heat conductivity of material A and B results in a
strongly inhomogeneous potential and temperature profile. A big part
of the temperature drop is located across material B, which is
characterized by $\alpha_B > \alpha_A$, hence the effective Seebeck
coefficient is enhanced close to the percolation threshold (see
fig.~\ref{fig:effective_cond_comp}).

The overestimated Seebeck coefficient $\alpha_\eff'$ in turn affects
the heat conductivity shown in fig.~\ref{fig:effective_cond_comp}
(right) leading to a an overestimation of $\kappa_\eff'$ by a factor
larger than $3$. Taking $\alpha_\eff$ determined by
eq.~(\ref{eq:seebeck_comp}) to derive $\kappa_\eff'$ with
eq.~(\ref{eq:harman_heat_conductance}) (yellow squares) a
significantly better agreement with effective medium theory (black
line) is achieved.

\section{Summary}

A simple but powerful simulation model has been derived that describes
all thermoelectric responses according to the Onsager-de Groot-Callen
transport theory.  Although not discussed in the present work we would
like to emphasize, that the generalizations to arbitrary (including
three dimensional) lattices and the inclusion of charge dynamics
\cite{Hartner2012} are straight forward.

By means of this model we pointed out that the Harman method to
measure the transport coefficients experimentally has substantial
systematic errors, when applied to composite materials. In its usual
operation mode (weak imposed current, weak thermal coupling to the
environment) it always overestimates the absolute value of the Seebeck
coefficient and the figure of merit. We gave a numerical example,
where the Seebeck coefficient turned out to be wrong by more than a
factor of 2 (fig.~\ref{fig:effective_cond_comp}, $f\approx 0.5$).

In order to explain this effect, we calculated it analytically for a
superlattice of alternating layers of two different materials. The
reason is that the Peltier heating/cooling at the interfaces enlarges
the temperature gradient in the layers with the larger absolute value
of the Seebeck coefficient, and reduces it in the other layers. This
self organized correlation between Seebeck coefficient and temperature
gradient persists, when the current is switched off. The temperature
profile acts as a memory of the previous current and affects the
thermal voltage, from which the Seebeck coefficient is inferred in the
Harman method. The correct Seebeck coefficient would be measured under
open circuit conditions for a different internal temperature profile
that does not reflect any previous Peltier heating/cooling. The
difference between the two measurements of the Seebeck coefficient
could be calculated analytically for a superlattice and is given in
eq.~(\ref{eq:seebeck_harman_analytic}). It can be used to correct the
Harman measurement.

Overestimating the Seebeck coefficient implies an overestimation of
the heat conductivity in the Harman method, as well. However, for
small currents we determined an expression,
eq.~(\ref{eq:heat_cond_harman_analytic}), which represents the correct
open circuit heat conductivity using quantities available during a
Harman measurement.

Random composites are only accessible by simulations. In this paper we
discussed a material composed of metal and semiconductor particles or
domains (fig.~\ref{fig:dist_comp}).  Below the percolation threshold
of the metal-like phase we confirmed that the Harman method
overestimates $\alpha$ and $\kappa$.

In summary, the application of the Harman method for inhomogeneous
media is tricky. However, for segmented thermoelectrics including
superlattices results from the Harman method can be corrected.

\section{Acknowledgment}
We thank R. Chavez, G. Schierning and R. Schmechel for valuable
discussions. Financial support by the German Research Foundation (DFG)
within the Priority Program on nanoscaled thermoelectric materials,
SPP 1386 is gratefully acknowledged.

\bibliography{bibliography}

\end{document}